\documentstyle[preprint,aps]{revtex} 
\tightenlines 
\begin{document} 
\draft 
\title{ 
Magnon-Paramagnon Effective Theory of 
Itinerant Ferromagnets 
         } 
\author{Naoum Karchev\cite{byline}} 
\address{ 
Department of Physics, University of Sofia, 1126 Sofia, Bulgaria 
      } 
\maketitle 
\begin{abstract} 
The present work is devoted to the derivation of an effective magnon-paramagnon 
theory starting from a microscopic lattice model of ferromagnetic metals. 
For some values of the microscopic 
parameters it reproduces the Heisenberg theory of localized spins. 
For small magnetization the effective model describes the physics 
of weak ferromagnets in accordance with the experimental results. 
It is written in a way which keeps $O(3)$ symmetry manifest,and describes both 
the order and disordered phases of the system. 
 
Analytical expression for the Curie temperature,which takes the magnon 
fluctuations into account exactly, is obtained.  
For weak ferromagnets $T_c$ is well below the Stoner's critical temperature 
and the critical temperature obtained within 
Moriya's theory. 
 
\end{abstract} 
{\bf I Introduction} 
 
The Heisenberg model of ferromagnetism, based on the exchange interaction 
of localized electrons gives an explanation of many properties of 
non-conducting magnetic systems at both low and high temperatures. However the 
development of a satisfactory theory of ferromagnetic metals has ran into 
difficulties. 
 
Theories of weak ferromagnetic metals based on the Landau Fermi liquid theory 
have been developed by several theorists \cite{mp1,mp2,mp3,mp4}. The 
spectrum of the spin excitations has been found. It consists of spin 
fluctuations of paramagnon type and a transverse spin-wave branch. 
 
Murata and Doniach \cite{mp6} have proposed a phenomenological mode-mode 
coupling theory to describe the temperature dependence of the quantities for 
a weak ferromagnet. However, they started from a classical Hamiltonian and 
ignored the quantum effects, which are important because of the low 
temperature nature of the weak ferromagnet. 
 
An alternative approach to magnetic phase transition in Fermi systems  
has been developed by Moriya and  Kawabata \cite{mp10,mp10a}. It is a 
self-consistent one loop approximation which accounts for the spin 
fluctuations. 
 
The nonlinear effects of spin fluctuations are treated and in \cite{mp10b}, 
using a self-consistent 
rotationally invariant Hartree approximation. 
 
New results concerning magnetic phase diagram in Hubbard-type 
model have been obtained within the dynamical mean-field theory 
\cite{mp11,mp12}. 
 
Perhaps the most striking feature of the itinerant ferromagnets is 
the quantum transition to paramagnet. In the paper \cite{mp7}, 
Hertz derives an effective paramagnon theory of a 
paramagnet to ferromagnet quantum phase transition. He analyzed 
the effective model by means of a renormalization group method 
that generalizes Wilson's treatment of classical phase transition 
and concludes that the critical behavior of itinerant ferromagnets 
in dimensions larger then one are described by a mean-field fixed 
point. Hertz's work has been reexamined in Ref.\cite{mp8}. 
The results of this paper support those of Hertz in 3D case, 
but in many aspects differ from 
them in two dimensions. 
 
More realistic model of ferromagnetic metals is considered in Refs 
\cite{mp9,mp17}. It contains particle-hole spin-triplet 
interaction that causes ferromagnetism, as well as particle-hole 
spin-singlet and particle-particle interactions. The spin-triplet 
interaction is decoupled by introducing a vector field whose 
average is proportional to the magnetization, and performing a 
Hubbard-Stratanovich transformation. Then the fermionic degrees of 
freedom are integrated out accounting for the rest part of 
interaction by means of perturbation theory. The resulting 
effective field theory is nonlocal. It contains an effective 
long-range interaction between the order parameter fluctuations. 
The analysis in Refs \cite{mp9,mp17} is restricted to power 
counting arguments at tree level. It shows that the critical 
behavior is governed by Gaussian fixed point and that all 
non-Gaussian terms are irrelevant in renormalization group sense. 
Logarithmic corrections to power-low scaling are obtained in $D=3$ 
case. For $D<3$ the deviations from the mean-field theory results 
depend on $D$. 
 
There is a considerable interest in the finite temperature properties of weak 
itinerant ferromagnets. It was obtained in \cite{mp4}, that in the region below 
but close to the Curie temperature, the paramagnon contribution is dominant. 
Experimentally, the transition in the weak ferromagnet $MnSi$ has been 
investigated at different Curie temperatures by applying hydrostatic pressure 
\cite{mp16}. The transition at high temperatures was found to be of second 
order, while at lower transition temperatures it is of first order. In the 
first order regime the transition temperature was found to scale with pressure 
as $T_c\sim (p_c\,-\,p)^{\frac 12}$. The scaling law is explained by a 
mean-field analysis, assuming a dynamical exponent $z=3$. But in Hertz's 
theory \cite{mp7}, the dynamical critical exponent is equal to three 
due to paramagnon excitations. Hence, in order to 
correctly explain the characteristic features of itinerant ferromagnets it 
should be important to work with an effective theory that keeps explicitly 
the magnon as well as paramagnon excitations. 
 
The present work is devoted to the derivation of an effective 
magnon-paramagnon theory starting from a microscopic lattice model 
of ferromagnetic metals. This is an effective theory of 
ferromagnets, and for some values of microscopic parameters it 
reproduces the Heisenberg theory of localized spins. On the other 
hand, for small magnetization the effective model describes the 
physics of weak ferromagnets in accordance with the experimental 
results. Thus, the effective magnon-paramagnon theory interpolates 
between theories of localized and itinerant electrons. The zero 
temperature dimensionless magnetization per lattice site $m$ is 
introduced to describe this interpolation. It parameterizes the 
ground state of the system, and moving $m$ we change the ground 
state and respectively the fluctuations above it. When $m$ is 
maximal ($m=\frac 12$) all the lattice sites are occupied by one 
electron with spin up in the ground state, and the only relevant 
excitations are magnons. The effective theory is the Heisenberg 
theory of localized spins. When the magnetization is smaller, i.e. 
some of the lattice sites are doubly occupied or empty in the 
ground state, the spectrum consists of paramagnon and magnon 
excitations and the effective theory is a  "spin $m$" Heisenberg 
theory coupled to paramagnon fluctuations. Decreasing the 
magnetization results in the changing the parameters of the spin 
fluctuations, and close to quantum critical point $m=0$, the 
paramagnon becomes important due to the singularity in the 
paramagnon propagator. In the quantum paramagnetic phase $(m=0)$, 
the magnon excitations disappear from the spin spectrum and one 
obtains Hertz's effective model. 
 
The ferromagnetic order parameter is a vector $\vec M$ field. 
The  transverse spin fluctuations (magnons) are described by $M_1+iM_2$ 
($M_1-iM_2$) and the longitudinal  fluctuations (paramagnons) by  
$M_3-<M_3>$. Alternatively the vector field can be written as a product of 
its amplitude $\rho=\sqrt{M_1^2+M_2^2+M_3^2}$ and an unit vector $\vec n$, 
$\vec M=\rho\vec n$. In the ferromagnetic phase one sets $M_3=<M_3>+\varphi$ 
and in linear (spin-wave) approximation obtains $\rho=<M_3>+\varphi$. 
It is evident now, that the fluctuations of the $\rho$ field,  
$\rho-<M_3>$ are exactly the paramagnon excitations in a formalism 
which keeps $0(3)$ symmetry manifest. One can write the 
effective theory in terms of $\vec M$-vector components, or in terms of $\rho$ 
and an unit vector $\vec n$. In ferromagnetic phase, in spin-wave 
approximation the effective theories coincide. I use the parameterization in 
terms of unite vector and spin singlet amplitude. Then, the  effective action 
keeps the $O(3)$ symmetry manifest, and describes both the order and disorder 
phases of the system. 
Above Curie temperature the spectrum consists of spin 
singlet fluctuations of paramagnon type and spin-$\frac 12$ spinon 
fluctuations. The spinon has a gap, but near the critical temperature it  
approaches zero and spin-$\frac 12$ fluctuations as well as paramagnon are 
essential describing the thermal phase transition of itinerant ferromagnets. 
 
Analytical expression for the Curie temperature,which takes the magnon 
fluctuations into account exactly, is obtained. For weak 
ferromagnets the critical temperature scales like $T_c\sim m^{\frac 53}$. 
It is well below the Stoner's critical temperature $T_c\sim m$ 
and the critical temperature obtained within 
Moriya's theory $T_c\sim m^{\frac 32}$ \cite{mp19}. 
 
Scaling arguments based on the Hert's renormalization group technique allow to 
obtain a relation between spin stiffness and the Stoner's enhancement factor. 
The relation is a consequence of the fact that quantum critical exponent is 
equal to three. Making use of the relation one can obtain the contribution of 
magnon scattering to the Stoner's enhancement factor. 
 
To derive the effective model a Schwinger-bosons-slave-fermions 
representation of the operators is used. The salient point is that 
within this approach the local Coulomb repulsion is treated 
exactly. As a result, the constants in the effective theory are 
finite and well defined for all values of the magnetization $m$. 
 
The paper is organized as follows. In Sec.II an effective magnon paramagnon 
theory of ferromagnetic metals is obtained. The Curie temperature is 
calculated and written in terms of magnetization and spin stiffness constant. 
In Sec III, Hertz's renormalization group in tree approximation is extended 
to include the magnon fluctuations. Section IV is devoted to the concluding 
remarks. 
 
\ \\

{\bf II Magnon-paramagnon effective model} 
 
The simplest model of ferromagnetic metals is determined by the 
Hamiltonian \cite{mp18} 
\FL 
\begin{equation} 
\hat H = -t \sum\limits_{<i,j>,\sigma} 
(\hat{c}^+_{i\sigma} \hat{c}_{j\sigma} + {\rm h.c.}) 
-J\sum\limits_{<i,j>} 
{\hat{\vec {S}}}_i\cdot {\hat{\vec {S}}}_j  
+ U \sum\limits_{i} \hat{n}_{i\uparrow} \hat{n}_{i\downarrow} - 
\mu\sum\limits_i \hat n_i 
\label{mp1} 
\end{equation} 
Here $\hat{c}^+_{i\sigma}$ and $\hat{c}^{\phantom +}_{j\sigma}$ are creation 
and annihilation operators for electrons, 
$\hat{n}_{i \sigma} = \hat{c}^+_{i \sigma}\hat{c}^{\phantom +}_{i \sigma}$ 
and $\hat{n}_{i} = \sum_{\sigma} \hat{n}_{i\sigma}$ are density operators, 
and ${\hat{\vec{S}}}_{i}=1/2 \sum\limits_{\sigma\sigma'} \hat{c}^+_{i\sigma} 
{\vec {\tau}}_{\sigma\sigma'} \hat{c}^{\phantom +}_{i\sigma'}$, 
where ${\vec {\tau}}$ denotes the vector of Pauli matrices, are spin operators. 
The sums are over all sites of a three-dimensional cubic lattice, 
$<i,j>$ denotes the sum over the nearest neighbors, and $\mu$ is the chemical 
potential. 
In (\ref{mp1}) the $J$-term corresponds to direct Heisenberg exchange which is 
generically ferromagnetic ($J>0$) in nature. 
 
In terms of Schwinger-bosons ($\hat \varphi_{i,\sigma},\hat 
\varphi_{i,\sigma}^{\dagger}$) and slave-fermions ($\hat h_i,\hat 
h_i^{\dagger},\hat d_i,\hat d_i^{\dagger}$) the operators have the 
following representation: 
\FL 
\begin{eqnarray} 
& & \hat {c}_{i\uparrow} = \hat h_i^{\dagger}\hat \varphi _{i1}+ 
\hat \varphi_{i2}^{\dagger}\hat d_i, 
\qquad 
\hat {c}_{i\downarrow} = \hat h_i^{\dagger}\hat \varphi _{i2}- 
\hat \varphi_{i1}^{\dagger}\hat d_i, 
\qquad 
\hat n_i = 1 - \hat h^+_i\hat h_i + \hat d^+_i\hat d_i, 
\nonumber \\ 
& & \hat {c}_{i\uparrow}^{\dagger}\hat {c}_{i\uparrow} 
\hat {c}_{i\downarrow}^{\dagger}\hat {c}_{i\downarrow} = 
\hat d_i^{\dagger}\hat d_i, 
\qquad 
\hat{\vec{S}}_{i}=\frac 12 \sum\limits_{\sigma\sigma'} \hat\varphi^+_{i\sigma} 
{\vec {\tau}}_{\sigma\sigma'} \hat\varphi_{i\sigma'},\qquad 
\hat \varphi_{i\sigma}^{\dagger}\hat \varphi_{i\sigma}\, 
+\,\hat d_i^{\dagger}\hat d_i\,+\,\hat h_i^{\dagger}\hat h_i\,=\,1 
\label{mp2} 
\end{eqnarray} 
The partition function can be written as a path integral over the complex 
functions of the Matsubara time $\tau$\,\, $\varphi_{i\sigma}(\tau)\, 
\left(\bar\varphi_{i\sigma}(\tau)\right)$ and Grassmann functions 
$h_i(\tau)\,\left(\bar h_i(\tau)\right)$ and $d_i(\tau)\, 
\left(\bar d_i(\tau)\right)$. 
\FL 
\begin{equation} 
{\cal Z}(\beta)\,=\,\int\,D\mu\left(\bar\varphi,\varphi,\bar h,h,\bar d,d,\right) 
e^{-S}. 
\label{mp4} 
\end{equation} 
The action is given by the expression 
\FL 
\begin{equation} 
S\,=\,\int\limits^{\beta}_0 d\tau\left[\sum\limits_i\left(\bar\varphi_{i\sigma} 
(\tau) 
\dot\varphi_{i\sigma}(\tau)\,+\,\bar h_i(\tau)\dot h_i(\tau)\,+ 
\,\bar d_i(\tau)\dot d_i(\tau)\right)\,+\, 
H\left(\bar\varphi,\varphi,\bar h,h,\bar d,d\right)\,\right], 
\label{mp5} 
\end{equation} 
where $\beta$ is the inverse temperature and the Hamiltonian is obtained 
from Eqs.(\ref{mp1}) and (\ref{mp2}) 
replacing the operators with the functions. In terms of Schwinger 
bosons and slave-fermions the theory is $U(1)$ gauge invariant, and the measure 
includes $\delta$ functions that enforce the constraint  and the 
gauge-fixing condition 
\FL 
\begin{eqnarray} 
D\mu\left(\bar\varphi,\varphi,\bar h,h,\bar d,d\right) & = & 
 \prod\limits_{i,\tau,\sigma}\frac {D\bar\varphi_{i\sigma}(\tau) 
D\varphi_{i\sigma}(\tau)}{2\pi i}\prod\limits_{i\tau}D\bar h_i(\tau) 
D h_i(\tau)D\bar d_i(\tau)D d_i(\tau) \nonumber \\ 
& & \prod\limits_{i\tau}\delta\left(\bar\varphi_{i\sigma}(\tau) 
\varphi_{i\sigma}(\tau)\,+\,\bar h_i(\tau) h_i(\tau)\,+\, 
\bar d_i(\tau) d_i(\tau)\,-\,1\right) 
\prod\limits_{i\tau}\delta\left(g.f\right). 
\label{mp6} 
\end{eqnarray} 
 
I make a change of variables, introducing new Bose fields $f_{i\sigma}(\tau) 
\left(\bar f_{i\sigma}(\tau)\right)$ Ref.\cite{mp13} 
\FL 
\begin{eqnarray} 
f_{i\sigma}(\tau) & = & \varphi_{i\sigma}(\tau) 
\left(1-\bar h_i(\tau)h_i(\tau)-\bar d_i(\tau)d_i(\tau)\right)^ 
{-\frac 12},\nonumber \\ 
\bar f_{i\sigma}(\tau) & = &\bar \varphi_{i\sigma}(\tau) 
\left(1-\bar h_i(\tau)h_i(\tau)-\bar d_i(\tau)d_i(\tau)\right)^ 
{-\frac 12}, 
\label{mp7} 
\end{eqnarray} 
where the new fields satisfy the constraint 
\FL 
\begin{equation} 
\bar f_{i\sigma}(\tau)f_{i\sigma}(\tau)\,=\,1. 
\label{mp9} 
\end{equation} 
In terms of the new fields the spin vector and the 
action have the form 
\FL 
\begin{equation} 
\vec{S}_{i}(\tau)=\frac 12 \sum\limits_{\sigma\sigma'} f^+_{i\sigma}(\tau) 
{\vec {\tau}}_{\sigma\sigma'} f_{i\sigma'}(\tau) 
\left(1-\bar h_i(\tau)h_i(\tau)-\bar d_i(\tau)d_i(\tau)\right) 
\label{mp10a} 
\end{equation} 
\FL 
\begin{eqnarray} 
S & = & \int\limits^{\beta}_0 d\tau \left\{\sum\limits_i 
\left[\bar f_{i\sigma}(\tau)\dot f_{i\sigma}(\tau)\,+\, 
\bar h_i(\tau)\left(\frac {\partial}{\partial\tau}-\bar f_{i\sigma}(\tau) 
\dot f_{i\sigma}(\tau) 
\right)h_i(\tau) 
\right.\right. \nonumber \\ 
& & \left.\left.+\,\bar d_i(\tau)\left(\frac {\partial}{\partial\tau}- 
\bar f_{i\sigma}(\tau)\dot f_{i\sigma}(\tau) 
\right)d_i(\tau)\right]\,+\,H\left(\bar f,f,\bar h,h,\bar d,d\right)\right\}, 
\label{mp10} 
\end{eqnarray} 
where $H\left(\bar f,f,\bar h,h,\bar d,d\right)$ is  the Hamiltonian 
\FL 
\begin{eqnarray} 
H & = & -t\sum\limits_{<i,j>}\left[ 
\bar d_i d_j \bar f_{j\sigma} f_{i\sigma}\,+ 
\,\bar d_j d_i \bar f_{i\sigma} f_{j\sigma}\,- 
\,\bar h_i h_j \bar f_{j\sigma} f_{i\sigma}\,- 
\,\bar h_j h_i \bar f_{i\sigma} f_{j\sigma}\right. \nonumber \\ 
& &\left.+\,\left(h_id_j\,-\,h_jd_i\right)\left(\bar f_{i1}\bar f_{j2}\,-\, 
\bar f_{i2}\bar f_{j1}\right) 
\,+\,\left(\bar d_j\bar h_i\,-\,\bar d_i\bar h_j\right)\left(f_{i1} f_{j2}\,-\, 
f_{i2} f_{j1}\right)\right] \nonumber \\ 
& & \times \left(1\,-\,\bar h_i h_i\,-\,\bar d_i d_i \right)^{\frac 12} 
\left(1-\bar h_j h_j\,-\,\bar d_j d_j\right)^{\frac 12} \nonumber \\ 
& & +\,\frac J2 \sum\limits_{<i,j>}\left[\bar h_i h_i\,+\,\bar d_i d_i\,+\, 
\bar h_j h_j\,+\,\bar d_j d_j\right]\, 
-\,\frac J2 \sum\limits_{<i,j>}\left(\bar h_i h_i\,+\bar d_i d_i\right) 
\left(\bar h_j h_j\,+\bar d_j d_j\right) \nonumber \\ 
& & +\frac J8\sum\limits_{<i,j>} 
\left(\vec n_j\,-\,\vec n_i\right)^2 
\left(1-\bar h_i h_i\,-\,\bar d_i d_i \right)^{\frac 12} 
\left(1-\bar h_j h_j\,-\,\bar d_j d_j \right)^{\frac 12} \nonumber \\ 
& & +U\sum\limits_{i}\bar d_i d_i\,-\,\mu\sum\limits_{i}\left(1\,- 
\,\bar h_i h_i\,+\,\bar d_i d_i\right) 
\label{mp11} 
\end{eqnarray} 
In Eq.(\ref{mp11})  
$\vec n_i\,=\,\sum\limits_{\sigma\sigma'}\bar 
f_{i\sigma}\vec\tau_{\sigma,\sigma'}f_{i\sigma'}$  is an unit vector. 
The Eq.({\ref{mp10a}}) describes in an $O(3)$ covariant way the spin. 
When the lattice site is empty or doubly occupied the spin 
vector is zero. When the lattice site is occupied by one electron 
the unit vector  $\vec n_i$  identifies the local orientation. 
One can consider 
 the first two components $n_{i1}$ and $n_{i2}$ as 
independent, and then $n_{i3}\,=\,\sqrt{1\,-\,n_{i1}^2\,-\,n_{i2}^2}$. 
In the leading order of the fields, the spin vector has the form      
\FL 
\begin{equation} 
S_{i1}\simeq \frac 12 n_{i1},  
\qquad 
S_{i2}\simeq \frac 12 n_{i2}, 
\qquad 
S_{i3}\,-\,\frac 12 \simeq -\frac 12 \left(\bar h_i h_i\,+\,\bar d_i 
d_i\right).  
\label{mp11a} 
\end{equation}  
The last equation shows that the longitudinal spin fluctuations are associated with 
the collective fields $(\bar h_i h_i\,+\,\bar d_i d_i)$.  
 
To avoid 
misunderstandings, it is important to point out that the charge-waves are 
associated with the collective field $(\bar d_i d_i\,-\,\bar h_i h_i)$ (see 
the representation of the electron number operator (\ref{mp2})).

To formulate a mean-field theory I drop the terms of order equal or higher 
then six in the Hamiltonian Eq.(\ref{mp11}). 
It is convenient to replace the term 
$\frac J2 \sum\limits_{<i,j>}\left(\bar h_i h_i\,+\bar d_i d_i\right) 
\left(\bar h_j h_j\,+\bar d_j d_j\right)$ in Eq.(\ref{mp11}) 
with the local one 
$\frac {3J}{2} \sum\limits_{i}\left(\bar h_i h_i\,+\bar d_i d_i\right)^2$. 
The difference is of higher order of derivatives, and I'll drop it. 
Then one can decouple this term, by means of the Hubbard-Stratanovich 
transformation, 
introducing a real, spin-singlet and gauge invariant field. 
\FL 
\begin{eqnarray} 
& & e^{\frac {3J}{2}\int\limits^{\beta}_0 d\tau\sum\limits_i 
\left(\bar h_i(\tau) h_i(\tau)\,+\,\bar d_i(\tau) d_i(\tau)\right)^2}\nonumber \\ 
& & =\int\prod\limits_{i\tau} DS_i(\tau) 
e^{-\int\limits^{\beta}_0 d\tau\sum\limits_i 
\left[\frac {3J}{2}S_i(\tau) S_i(\tau)\,-\,3J \left(\bar h_i(\tau) h_i(\tau)\,+ 
\bar d_i(\tau) d_i(\tau)\right)S_i(\tau)\right]} 
\label{mp13} 
\end{eqnarray} 
Now, the action is quadratic 
with respect to the fermions and one can integrate them out. 
The resulting action depends on the spinons and the real field $S_i$. It has  
a minimum at the point $S_i=s_0, f_{i\sigma}=f_{\sigma}$, and the stationary  
condition is  
\FL  
\begin{equation}  
s_0\,=\,<\bar h_i h_i\,+\,\bar d_i d_i>  
\label{mp14}  
\end{equation}  
Expanding the effective action around this point one obtains the effective 
model.  
 
To improve the calculations I account for the terms of order six and higher, 
replacing the collective field $\bar h_i h_i\,+\,\bar d_i d_i$ in these 
terms by its mean-field value from Eq.({\ref{mp14}). The new Hamiltonian 
depends on the fields $\bar f_{i\sigma}(\tau)\,,\,f_{i\sigma}(\tau)\,,\, 
2\varphi_i(\tau)=s_0-S_i(\tau)$, and is quadratic with respect to the 
fermions  
\FL  
\begin{eqnarray}  
H & = & -t\sum\limits_{<i,j>}\left[\bar d_i d_j\,+ \,\bar d_j d_i\, 
- \,\bar h_i h_j \,- \,\bar h_j h_i\right]\,+\, 
\left(6mJ+U\,-\mu\right)\sum\limits_{i}\bar d_i d_i\,+\, 
\left(6mJ+\mu\right)\sum\limits_{i}\bar h_i h_i \nonumber \\ 
& & -2mt\sum\limits_{<i,j>}\left[  
\left(\bar d_i d_j\,-\,\bar h_i h_j\right)\left(\bar f_{j\sigma} 
f_{i\sigma}\,-\,1\right)\,+\,\ 
\left(\bar d_j d_i\,-\,\bar h_j h_i\right)\left(\bar f_{i\sigma} 
f_{j\sigma}\,-\,1\right)\right. \nonumber \\  
& & \left.+\,\left(h_id_j\,-\,h_jd_i\right)\left(\bar f_{i1}\bar f_{j2}\,-\,  
\bar f_{i2}\bar f_{j1}\right)  
\,+\,\left(\bar d_j\bar h_i\,-\,\bar d_i\bar h_j\right)\left(f_{i1} f_{j2}\,-\,  
f_{i2} f_{j1}\right)\right] \nonumber \\  
& & +6J\sum\limits_i\varphi_i \left(\bar h_i h_i\,+ \bar d_i d_i\right)  
+\frac {(2m)^2J}8\sum\limits_{<i,j>}  
\left(\vec n_j\,-\,\vec n_i\right)^2    
\label{mp13a}  
\end{eqnarray}  
where $m=\frac 12(1-s_0)$. 
Integrating out the fermions one obtains the action of the effective theory. 
 
To get intuition how the effective action looks like, it is important to stress 
that the spinon fields contribute the action through the fields 
$\bar f_{i\sigma}(\tau)\dot f_{i\sigma}(\tau)$,\,\, 
$\bar f_{i\sigma}(\tau)\left( 
f_{j\sigma}(\tau)-f_{i\sigma}(\tau)\right)$,\,\, 
$\left(f_{i1}(\tau)f_{j2}(\tau)\,-\,f_{j1}(\tau)f_{i2}(\tau)\right)$\,\,and 
$\left(\vec n_j\,-\,\vec n_i\right)^2$ (see Eq.\ref{mp13a}) 
In the continuum limit they have the form 
$\bar f_{\sigma}\partial_{\mu}f_{\sigma}$\,$(\mu=\tau,x,y,z)$,\,\, 
$\left(f_1\partial_{\nu}f_2-f_2\partial_{\nu}f_1\right)$\,\, 
$(\nu=x,y,z)$ and 
$\partial_{\nu}\vec n\cdot\partial_{\nu} \vec n$. 
The four-vector\, $A_{\mu}=i\bar f_{\sigma}\partial_{\mu}f_{\sigma}$\, transforms 
as an\, $U(1)$\, gauge field,\, 
$\left(f_1\partial_{\nu}f_2-f_2\partial_{\nu}f_1\right)$\, 
as a charge-two complex field and 
\,$\partial_{\nu}\vec n\cdot\partial_{\nu} \vec n$ is a gauge invariant. 
Hence, the simplest gauge invariant and spin- 
singlet contributions of the first two fields have the form 
$\left(\bar f_1\partial_{\nu}\bar f_2-\bar f_2\partial_{\nu}\bar f_1 
\right)\left(f_1\partial_{\nu}f_2-f_2\partial_{\nu}f_1\right)$ and 
$\left(\partial_{{\mu}_1}A_{{\mu}_2}-\partial_{{\mu}_2}A_{{\mu}_1}\right) 
\left(\partial_{{\mu}_1}A_{{\mu}_2}-\partial_{{\mu}_2}A_{{\mu}_1}\right)$. 
It is not difficult to check that 
\FL 
\begin{equation} 
\left(\bar f_1\partial_{\nu}\bar f_2-\bar f_2\partial_{\nu}\bar f_1 
\right)\left(f_1\partial_{\nu}f_2-f_2\partial_{\nu}f_1\right)\,=\, 
\frac 14\partial_{\nu}\vec n\cdot\partial_{\nu} \vec n. 
\label{mp15} 
\end{equation} 
The term $\left(\partial_{{\mu}_1}A_{{\mu}_2}-\partial_{{\mu}_2} 
A_{{\mu}_1}\right) 
\left(\partial_{{\mu}_1}A_{{\mu}_2}-\partial_{{\mu}_2}A_{{\mu}_1}\right)$ 
is of the same order as $\left(\partial_{\nu}\vec 
n\cdot\partial_{\nu} \vec n\right)^2$ and I'll ignore it. To 
obtain the effective theory it is convenient to set the gauge 
field $A_{\mu}=i\bar f_{\sigma}\partial_{\mu}f_{\sigma}$ equal to 
zero, and to account for the contribution of the complex field 
$\left(f_1\partial_{\nu}f_2-f_2\partial_{\nu}f_1\right)$, the real 
field $\varphi$, and $\partial_{\nu}\vec n\cdot\partial_{\nu} \vec 
n$. An important exclusion is the  $\bar 
f_{\sigma}\partial_{\tau}f_{\sigma}$ field which contributes the 
action linearly. 
 
Expanding the effective functional around the mean-field point and keeping 
only the first three terms, one can write the effective 
action in the form 
  
\FL 
\begin{equation} 
S_{\text eff}\,=\,S_{\text H}\,+\,S_{\text p}\,+\,S_{\text int}, 
\label{mp16} 
\end{equation} 
 
$S_{\text H}$ is the action of the Heisenberg theory of localized spins. 
In continuum limit it has the form 
\FL 
\begin{equation} 
S_{H}\,=\,\int d\tau d^3\vec r\left[2m\bar f_{\sigma}(\tau,\vec r)\dot 
f_{\sigma}(\tau,\vec r)\,+\,\frac {m^2J_{\text r}}{2}\sum\limits_{\nu=1}^3 
\partial_{\nu}\vec n(\tau,\vec r)\cdot\partial_{\nu}\vec n(\tau,\vec r) 
\right]. 
\label{mp17} 
\end{equation} 
In Eq.(\ref{mp17}),\, $m=\frac 12(1-s_0)$, and $s_0$ comes from 
"tadpole" diagrams with one $h$ or $d$ line. 
The renormalized exchange coupling constant has the following representation in 
terms of 
 microscopic parameters and at zero temperature \cite{mp14} 
\FL 
\begin{eqnarray} 
J_r & = & J\,-\,\frac {4t^2}{12Jm+U}\, 
+ \,\frac 83 \frac {t^2}{12Jm+U}\frac 1N\sum\limits_k\left(\sum 
\limits_{\nu=1}^{3} \sin^2 k_{\nu}\right)\left(n_k^h+n_k^d\right). 
\label{mp18} 
\end{eqnarray} 
In Eq.(\ref{mp18}),\,  $n_k^d$ and $n_k^h$ are the 
occupation numbers for $d$ and $h$ fermions respectively. The first term 
is due to the direct Heisenberg exchange term in Eq.(\ref{mp13a}), 
and the other terms are due to Anderson's superexchange. The last two 
terms are obtained calculating the one-loop self-energy diagrams 
 of $h$ and 
$d$ fermions. The superexchange contribution to the exchange coupling 
constant goes to zero for small magnetization.Hence, near the quantum phase 
transition one can replace the renormalized coupling constant $J_r$ by bare 
one $J$.

$S_p$ is the contribution to the effective 
action of the paramagnon excitations 
  
\FL 
  
\begin{equation} 
S_p\,=\,\frac 12 \int\frac {d\omega}{2\pi}\frac 
{d^3p}{(2\pi)^3}\varphi(\omega,\vec p) 
 \left( r\,+\,a\frac 
{|\omega|}{p}\,+\,b p^2\right)\varphi(-\omega,-\vec p) 
 \label{mp23} 
\end{equation} 
where 
\FL 
\begin{equation} 
r\,=\,12J\left[1\,-\,3J\left(N(\epsilon_F^h)\,+\,N(\epsilon_F^d) 
\right)\right] 
\label{mp24} 
\end{equation} 
and the constants $a$ and $b$ are calculated in continuum limit. 
\FL 
\begin{equation} 
a\,=\,18\pi J^2\left(\frac {N(\epsilon_F^h)}{v_F^h}\,+ 
\,\frac {N(\epsilon_F^d)}{v_F^d}\right)\quad 
b\,=\,3 J^2\left (\frac {N(\epsilon_F^h)}{\left(k_F^h\right)^2}\,+ 
\,\frac {N(\epsilon_F^d)}{\left(k_F^d\right)^2}\right). 
\label{mp25} 
\end{equation} 
It is obtained from the Lindhard functions for $h$ and $d$ fermions in 
the limit when $p$ and $\frac {\omega}{p}$ are small. 
 
Finally, the spinon-paramagnon interaction has the form 
\FL 
\begin{equation} 
S_{\text int}\,=\,m^2\lambda\int d\tau d^3\vec r\, \varphi(\tau,\vec r) 
\left[\sum\limits_{\nu =1}^{3}\partial_{\nu}\vec n(\tau,\vec r)\cdot 
\partial_{\nu}\vec n(\tau,\vec r)\right] 
\label{mp26} 
\end{equation} 
where 
\FL 
\begin{equation} 
\lambda\,=\, 
\frac {16Jt^2}{(12Jm+U)^2}\frac 
1N\sum\limits_k\left(\sum\limits_{\nu=1}^{3} 
 \sin^2 
k_{\nu}\right)\left(1\,-\,n_k^h-n_k^d\right). 
 \label{mp27} 
\end{equation} 
The effective magnon-paramagnon coupling is obtained from triangular 
diagrams with two $h$ and one $d$ 
 lines or with two $d$ and one $h$ lines. 
 
To analyze the effective model, it is more convenient to rewrite it in terms of 
rescaled spinon fields  
\FL  
\begin{equation}  
\bar \zeta_{i\sigma}\,=\,\sqrt{2m}\bar f_{i\sigma},\qquad 
\zeta_{i\sigma}\,=\,\sqrt{2m}f_{i\sigma}. 
\label{mp90a} 
\end{equation} 
The new fields satisfy the constraint 
\FL 
\begin{equation} 
\bar \zeta_{i\sigma}\zeta_{i\sigma}\,=\,2m , 
\label{mp90b} 
\end{equation} 
and the action of the effective theory has the form 
\FL  
\begin{eqnarray}  
S_{\text eff} & = & \int d\tau d^3\vec r\left[\bar \zeta_{\sigma}(\tau,\vec 
r)\dot  \zeta_{\sigma}(\tau,\vec r)\,+\,\frac {J_{\text 
r}}{2}\sum\limits_{\nu=1}^3  \partial_{\nu}\vec M(\tau,\vec 
r)\cdot\partial_{\nu}\vec M(\tau,\vec r)\right. \nonumber \\ 
& & \left.+\,\frac {\lambda}{4} 
\varphi(\tau,\vec r)  \left[\sum\limits_{\nu 
=1}^{3}\partial_{\nu}\vec M(\tau,\vec r)\cdot  \partial_{\nu}\vec M(\tau,\vec 
r)\right]\right]\,+\,S_p, 
\label{mp90c} 
\end{eqnarray} 
where $\vec M$ is the spin vector 
\FL 
\begin{equation} 
\vec M\,=\,\frac 12 \bar \zeta_{\sigma}         
\vec\tau_{\sigma,\sigma'}\zeta_{\sigma'}, \qquad 
\vec M^2\,=\,m^2 
\label{mp90d} 
\end{equation} 
and $S_p$ is given by Eq.(\ref{mp23}). 
 
It follows from Eq.(\ref{mp10a}) that the dimensionless magnetization 
of the system, per 
 lattice site is defined by the equation, 
\FL 
\begin{equation}  
<S_i^3>=\frac 12 <n_i^3>\left(1-<\bar h_i h_i+ \bar d_i d_i>\right). 
\label{mp90e} 
\end{equation}  
At zero temperature $<n_i^3>=1$ and 
using the Eq.(\ref{mp14}) one obtains that $m$ is zero temperature 
dimensionless magnetization of the system per lattice site, $m=<S_i^3>$. The 
parameter 
 $m$ depends on the microscopic parameters of the theory and 
characterizes 
 the vacuum. If, in the vacuum state, every lattice site is 
occupied by one 
 electron with spin up, then $m=\frac 12\quad (s_0=0)$, the 
parameters $a$ and 
 $b$ 
 from Eq.(\ref{mp23}) are equal to zero and $r=\frac 
{3J}{2}$. In this case one 
 can integrate over the paramagnons and the 
resulting theory is the spin $\frac 12$ Heisenberg 
 theory of the localized 
spins. When, in the vacuum state, some of the sites are 
 doubly occupied 
($<\bar d_i d_i>\neq 0$) or empty $(<\bar h_i h_i>\neq 0)$, 
 then $m<\frac 
12$, the relevant excitations are the spinon and paramagnon 
 excitations and 
the effective theory is a  "spin $m$" Heisenberg  
theory coupled to paramagnon fluctuations defined by  
Eqs.(\ref{mp90b},\ref{mp90c},\ref{mp90d}). The system approaches the 
quantum critical 
 point when $m\rightarrow 0\quad (s_0\rightarrow 1)$. One can 
see directly, from 
 the stationary condition (\ref{mp14}), that $r(m)$ 
approaches zero when 
 $m\rightarrow 0$. Hence, the parameter $r$ measures the 
distance from the quantum critical point. In quantum paramagnetic phase 
$(m=0)$, the spinon excitations disappear from the spin spectrum (see 
Eqs.(\ref{mp90b},\ref{mp90d})) and one obtains Hertz's effective model. One 
can add a four-paramagnon term , calculating one-loop diagrams with four $h$ 
or $d$ fermion lines, but I have dropped it motivated by the Hertz's result. 
 
The effective theory is $U(1)$ gauge invariant. Below the Curie temperature 
it is convenient to introduce explicitly the magnon excitations. To this end, 
I impose the gauge-fixing condition in the form 
 $arg \zeta_{i1}=0$. Then the 
constraint (\ref{mp90b}) can be solved by means of 
 the complex field 
$a_i(\tau)=\zeta_{i2}$ and $\zeta_{i1}=\sqrt {2m-\bar a_i(\tau) 
 a_i(\tau)}$. 
For the components of the spin vector 
$M^+\,=\,M_1\,+\,iM_2,\,\,M^-\,=\,M_1\,-\,iM_2,$ and $M_3$ one obtains the 
Holstein- 
 Primakoff representation:  
\FL 
\begin{eqnarray}  
M_i^+(\tau) & = & \sqrt {2m - \bar a_i(\tau)\,a_i(\tau)}\,\,a_i(\tau), \,\, 
M_i^-(\tau)\,=\,\bar a_i(\tau)\sqrt {2m - \bar a_i(\tau)\,a_i(\tau)}, 
\nonumber \\ 
M_i^3(\tau) & = & m\,-\,\bar a_i(\tau)\,a_i(\tau) 
\label{mp91} 
\end{eqnarray} 
The kinetic term in the 
action 
 and the measure are the same as the kinetic term and the measure in 
the theory 
 of Bose field. The only difference is that the complex fields are 
subject to 
 the condition $\bar a_i(\tau) a_i(\tau)\leq 1$. 
 
In the spin-wave theory one approximates $\sqrt {2m-\bar a_i(\tau)a_i(\tau)}$ 
and integrates over the whole complex plane. Then, the model is simplified and 
can be written in terms of magnon $a_i(\tau)\,\,(\bar a_i(\tau))$ and 
paramagnon $\varphi_i(\tau)$ fields 
\FL 
\begin{eqnarray} 
S_{\text eff} & = & \int \frac {d\omega}{2\pi}\frac {d^3p}{(2\pi)^3}\left[ 
\bar a(\omega,\vec p)\left (i\omega\,+\,\rho p^2\right) a(\omega,\vec p) 
\,+ 
\,\frac 12 \varphi(\omega,\vec p)\left(r\,+\,a\frac {|\omega|}{p}\,+\,b 
p^2\right) 
 \varphi(-\omega,-\vec p)\right] \nonumber \\ 
& + & \frac {m\lambda}{2} \int\prod\limits_{l=1}^{2}\frac {d{\omega}_l}{2\pi} 
\frac {d^3p_{l}} 
{(2\pi)^3}\left(\vec p_1\cdot\vec p_2\right)\bar a({\omega}_1,\vec p_1) 
a({\omega}_2,\vec p_2)\varphi({\omega}_1-{\omega}_2,\vec p_1-\vec p_2) 
\label{mp30} 
\end{eqnarray} 
where 
\FL 
\begin{equation} 
\rho\,=\,m\,J_r 
\label{mp30n} 
\end{equation} 
is the spin stiffness constant. 
 
Let's rewrite the spin vector Eq.(\ref{mp10a}) in terms of the vector $\vec M$ 
Eq.(\ref{mp90d}) and use the Holstein-  Primakoff representation 
Eq.(\ref{mp91}). Then the magnetization of the system per lattice site is given 
by the expression 
\FL 
\begin{equation}  
<S_i^3>\,=\,m\,-\,<\bar a_i a_i> 
\label{mp92} 
\end{equation}  
The equation for the critical temperature is $<S_i^3>\,=\,0$. Having in 
mind that the magnon dispersion is $\epsilon_a(p)\,=\,\rho p^2$ one 
obtains for the Curie temperature 
\FL 
\begin{equation} 
T_c\,=\,\kappa\,m^{\frac 23}\,\rho(m) 
\label{mp94} 
\end{equation} 
where the constant $\kappa$ can be written in terms of gamma $\Gamma(z)$ and 
Riemann $\zeta(z,q)$ functions $\kappa\,=\,(\Gamma(\frac 32)\zeta(\frac 
32,1)/4\pi^2)^{-\frac 23}$. In the spin-wave approximation the spin stiffness 
constant is given by Eq.(\ref{mp30n}), hence, when the system approaches 
quantum critical point ($m\rightarrow 0$), the critical temperature scales 
with magnetization like $T_c\sim m^{\frac 53}$. Recently, it was 
experimentally shown \cite{mp16}, that for weak ferromagnets the transition 
temperature  scales with pressure like  $T_c\sim (p_c\,-\,p)^{\frac 12}$.  To 
explain the experimental result one has to assume that magnetization scales 
with pressure like $m\sim (p_c\,-\,p)^{0,3}$. 
 
One can improve the equation for the Curie temperature replacing the zero 
temperature dimensionless magnetization $m$ by the finite temperature solution 
$m(T)$ of the mean-field equation Eq.(\ref{mp14}). 
\FL 
\begin{equation} 
T_c\,=\,\kappa\,m^{\frac 23}(T_c)\,\rho(m(T_c)) 
\label{mp94a} 
\end{equation} 
For conventional weak ferromagnets $m(T_c)\sim m$ and Eq.(\ref{mp94}) is an 
appropriate expression for Curie temperature. But for high $T_c$ weak 
ferromagnets $m(T_c)>m$ and the correct equation for the critical temperature 
is Eq.(\ref{mp94a}). 
 
It is important to stress that the equations for the critical temperature 
follow from the exact representations for the third component of the spin (\ref 
{mp10a},\ref {mp91},\ref {mp92}).Hence they take the magnon fluctuations into 
account exactly. 
 
In the spin-wave approximation the transverse components of the spin 
fields are 
 proportional to the magnon fields 
  
\FL 
  
\begin{equation} 
S_i^+(\tau)=\sqrt {2m} a_i(\tau),\,\,\,\,S_i^-(\tau)=\sqrt {2m} \bar 
a_i(\tau) 
 \label{mp30a} 
 \end{equation} 
and the field $\varphi_i(\tau)$ is exactly the paramagnon (longitudinal 
spin fluctuation) 
\FL 
\begin{equation} 
S_i^3(\tau)-<S_i^3>\,=\,\varphi_i(\tau). 
\label{mp30b} 
\end{equation} 
Hence, in Gaussian approximation, spin-spin 
correlation functions have the form 
\FL 
\begin{equation} 
D^{\text tr}(\omega,\vec p)\,=\,\frac {2m}{i\omega\,+\,\rho p^2}\,\,\,, 
\qquad\qquad 
D^{\text long}(\omega,\vec p)\,=\,\frac {1}{r\,+ 
a\frac {|\omega|}{p}\, 
+\,bp^2} 
\label{mp31} 
\end{equation} 
 
\ \\ 
 
{\bf III Scaling behavior}

Scaling arguments, based on the Hertz's 
 renormalization group technique 
allow to obtain a relation 
between the spin stiffness and the parameter $r$-the distance 
from the quantum critical point. 
 
To begin with, I introduce a cutoff and redefine the momenta 
introducing dimensionless one. RG construction starts with a definition of 
"soft" and "fast" modes, 
\FL   
\begin{eqnarray}  
\Psi(\omega,\vec p) & = & \Psi_{<}(\omega, \vec p)\,+\,\Psi_{>}(\omega, 
\vec p) \\ 
\Psi_{<}(\omega, \vec p) & = & \Psi(\omega, \vec p) \qquad {\text  for} 
\qquad |p|<e^{-l} \nonumber \\ 
\Psi_{>}(\omega, \vec p) & = & \Psi(\omega, \vec p) \qquad {\text  for} 
\qquad e^{-l}\leq |p| \leq 1 .\nonumber 
\label{mp95} 
\end{eqnarray} 
where $\Psi(\omega,\vec p)$ stands for magnons, $a(\omega,\vec 
p),\bar a(\omega,\vec p)$ and paramagnon $\varphi(\omega,\vec p)$ fields. 
The action Eq.(\ref{mp30}) depends on the "slow" and the "fast" modes. In the 
tree approximation one keeps only the "slow" modes 
\FL 
\begin{equation} 
S^{\text tree}(\bar a,a,\varphi)\,=\,S(\bar a_{<},a_{<},\varphi_{<}). 
\label{mp96} 
\end{equation} 
 
The next step is to change the variables, letting  
\FL 
\begin{equation} 
\vec p'\,=\,e^l\vec p, \qquad \omega'\,=\,e^{zl}\omega 
\label{mp97} 
\end{equation} 
The new momenta runs over the whole interval $0\leq |p'|\leq 1$, and in terms 
of the new variables the action has the form 
\FL  
\begin{eqnarray}  
S^{\text tree} & = & e^{-(z+3)l}\int \frac {d\omega'}{2\pi}\frac 
{d^3p'}{(2\pi)^3}\left[\bar a_{<}(e^{-zl}\omega',e^{-l}\vec p')\left 
(ie^{-zl}\omega'\,+\,mJ_r e^{-2l}p'^2\right) a_{<}(e^{-zl}\omega',e^{-l}\vec 
p')\right.  \nonumber \\  
& + &\left. \frac 12 \varphi_{<}(e^{-zl}\omega',e^{-l}\vec 
p')\left(r\,+\,e^{(-z+1)l}a\frac {|\omega'|}{p'}\,+\,e^{-2l}b 
p'^2\right)\varphi_{<}(-e^{-zl}\omega',-e^{-l}\vec p')\right] \nonumber \\   
& + & \Lambda e^{-2(z+4)}\int\prod\limits_{l=1}^{2}\frac {d{\omega'}_l}{2\pi}  
\frac {d^3p'_{l}}  {(2\pi)^3}\left(\vec p'_1\cdot\vec p'_2\right)\bar 
a_{<}({\omega'}_1,\vec p'_1)a_{<}({\omega'}_2,\vec 
p'_2)\varphi_{<}({\omega'}_1-{\omega'}_2,\vec p'_1-\vec p'_2),   
\label{mp98}  
\end{eqnarray} 
where I have used a short notation for the magnon-paramagnon coupling constant 
$\Lambda\,=\,\frac {m\lambda}{2}$. 
 
It is apparent that if one chooses $z\,=\,3$, the coefficient of $a\frac 
{|\omega'|}{p'}$, in the paramagnon quadratic term, is the same as the 
coefficient of $bp'^2$. Then rescaling the paramagnon $\varphi$, one can make 
the total coefficient of both of them unity. \FL 
\begin{equation} 
\varphi'(\omega',\vec p')\,=\,e^{-\frac 
{z+5}{2}l}\varphi_{<}(e^{-zl}\omega',e^{-l}\vec p') 
\label{mp99} 
\end{equation}  
To complete  
the RG transformation, one has to rescale the magnon  
fields too, to make the coefficient of $i\omega'$ unity, 
\FL 
\begin{equation} 
a_{<}(\omega',\vec p')\,=\,e^{-\frac 
{2z+3}{2}} a_{<}(e^{-zl}\omega',e^{-l}\vec p'), \qquad       
\bar a_{<}(\omega',\vec p')\,= 
\,e^{-\frac {2z+3}{2}}\bar a_{<}(e^{-zl}\omega',e^{-l}\vec p') 
\label{mp100} 
\end{equation}   
Then, for the transformed action one obtains 
\FL  
\begin{eqnarray}  
& S' & (\bar a',a',\varphi')\,=\,\int \frac {d\omega'}{2\pi}\frac 
{d^3p'}{(2\pi)^3}\left[  \bar a'(\omega',\vec p')\left (i\omega'\,+\,\rho' 
p'^2\right) a'(\omega',\vec p')\right. \nonumber \\ 
& + &\left. \frac 12 \varphi'(\omega',\vec p')\left(r'\,+\,a\frac 
{|\omega'|}{p'}\,+\,b p'^2\right)  \varphi'(-\omega',-\vec p',)\right] 
\nonumber \\  & + & \Lambda' \int\prod\limits_{l=1}^{2}\frac 
{d{\omega'}_l}{2\pi}  \frac {d^3p'_{l}}  
{(2\pi)^3}\left(\vec p'_1\cdot\vec p'_2\right)\bar a'({\omega'}_1,\vec p'_1)  
a'({\omega'}_2,\vec p'_2)\varphi'({\omega'}_1-{\omega'}_2,\vec p'_1-\vec p'_2)  
\label{mp101}  
\end{eqnarray}  
where $r',\rho'$ and $\Lambda'$ are the transformed parameters 
\FL  
\begin{eqnarray}  
r' & = & r(l)\,=\,re^{2l} \nonumber \\ 
\rho' & = & \rho(l)\,=\,\rho e^{(z-2)l}\,=\,\rho e^{l} \\ 
\Lambda' & = & \Lambda(l)\,=\,\Lambda e^{\frac {z-5}{2}l}\,=\,\Lambda e^{-l} 
\nonumber  
\label{mp102} 
\end{eqnarray} 
Excluding the scaling parameter $l$ one obtains the relations (RG invariants) 
\FL 
\begin{equation} 
\frac {r'}{\rho'^2}\,=\,\frac {r}{\rho^2}, \qquad 
r'\Lambda'^2\,=\,r\Lambda^2 
\label{mp103} 
\end{equation} 
 
The bare parameter $r$ scales with magnetization like $r\,=\,r_0m^2$, and the 
bare spin stiffness constant is proportional to the magnetization (see 
Eq.\ref{mp30}). Hence, for weak ferromagnets, one obtains 
\FL 
\begin{equation} 
r(m)\sim \rho^2(m) 
\label{mp104} 
\end{equation} 
 
 
The relation Eq.(\ref{mp104}) is a consequence of the fact that quantum 
critical exponent is equal to three ($z=3$). It holds even if the 
magnon-magnon interaction is introduced. Calculating the magnon contribution 
to the spin stiffness, one can use it to obtain the the magnon scattering 
corrections to the parameter $r$ (Stoner's enhancement factor) 
 
Making use of Eq.(\ref{mp104}) one can obtain that  
\FL 
\begin{equation} 
T_c\sim m^{\frac 23}\,T_s 
\label{mp106} 
\end{equation} 
where, $T_s\,=\,const\sqrt r\sim m$ is the Stoner's expression for the 
critical temperature (see Eq.\ref{mp24}). For weak ferromagnets $m<\frac 12$, 
and $T_c$ is well below the Stoner's critical temperature, as should 
be.Curie temperature obtained within Moriya's theory scales wit magnetization 
like $\sim m^{\frac32}$ \cite {mp19}.  
 
The result indicates that the spin-$m$ 
Heisenberg model coupled to paramagnon provides a good description of the 
itinerant ferromagnets.

 \ \\ 
 
{\bf IV Conclusions} 
 
An effective magnon-paramagnon theory of 
itinerant ferromagnets has been obtained. It is written in a way 
which keeps $O(3)$ symmetry manifest. 
 
The effective model differs from the models discussed in 
Refs.\cite{mp7,mp9,mp17} in two ways. First, it describes in unified way both 
the order 
 and disordered phases of the system. Altering the parameters, it 
interpolates between the Heisenberg theory of localized spins and 
 Hertz's 
theory of nearly ferromagnetic metals. In ferromagnetic phase the important 
spin fluctuations are transversal magnon fluctuations and longitudinal 
paramagnon fluctuations. In thermal paramagnetic phase (above Curie 
temperature) the spectrum consists of spin singlet fluctuations of paramagnon 
type and spin-$\frac 12$ spinon fluctuations. Well above the 
critical temperature the spinon has a large gap, and the physics of 
ferromagnetic metals is dominated by the paramagnon fluctuations. But just 
above $T_c$ the spinon's gap approaches zero \cite{mp20}, and the 
contribution of the spin-$\frac 12$ fluctuations is essential. Crossing the 
quantum critical point ($m=0$), the spinon excitations disappear from the 
spectrum and only the paramagnon survives in the quantum paramagnetic phase. 
 
Second, I have 
accounted for the on site Coulomb repulsion exactly. The Hubbard interaction 
is the strongest, and to drop it, as in Ref.\cite{mp7}, or to treat it 
perturbatively, as in Refs. \cite{mp9,mp17} is not adequate. As a result, I 
obtain that the vertices in the effective functional exist in the limit of zero 
frequencies and wavenumber, and that the constants are well defined for all 
values of the magnetization. 
 
The effective model Eqs.(\ref{mp90c},\ref{mp30}) enables to estimate  
the correctness of the Green's functions calculations \cite{mp4}, and 
the Moriya Kawabata approximation \cite{mp10}.Within these approaches, 
the spin stiffness constant is proportional to the magnetization. The same 
result follows from the effective model Eq.(\ref{mp30}) where the nonlinear 
magnon-magnon interaction is not considered. From spin-wave expansion follows 
that the magnon vertices are of order $(\frac 1m)^n$, and on the verge of 
ferromagnetism are relevant. One way to obtain the effects of magnon-magnon 
interaction is to use a renormalization group approach in the spin-wave 
theory, which allows for the analysis of systems with small spins \cite{mp21}. 
 
\ \\

\acknowledgments 
 
The author would like to thank K.Bedell and K.Blagoev for valuable 
discussions. This research is supported by the USA National 
Science Foundation Grant-DINT\#9876873.


\begin{references} 
%
\frenchspacing 
%
\bibitem[*]{byline} Electronic address: naoum@phys.uni-sofia.bg 
%
\bibitem{mp1} A. A. Abrikosov and I. E. Dzyloshinskii, Zh. Eksp. Teor. Fiz. 
{\bf 35}, 771 (1959) [Sov. Phys. JETP. {\bf8}, 535 (1959)]. 
%
\bibitem{mp2} T. V. Ramakrishnan, Solid State Commune. {\bf 14}, 449 (1974); 
Phys. Rev. B {\bf 10}, 4014 (1974). 
%
\bibitem{mp3} A. Kawabata, J. Phys. F {\bf 4}, 1477 (1974). 
%
\bibitem{mp4} I. E. Dzyaloshinskii and P. S. Kondratenko, Zh. Eksp. Teor. 
Fiz. 
 {\bf 70}, 1987 (1974) [Sov. Phys. JETP, {\bf 43}, 1036 (1975)]. 
%
\bibitem{mp6} K. Murata and S. Doniach, Phys. Rev. Lett. {\bf 29}, 285 (1972) 
%
\bibitem{mp10} T. Moriya and A. Kawabata, J. Phys. Soc. Japan {\bf34},639 
%
\bibitem{mp10a} The works are summarized in T. Moriya, {\it Spin 
Fluctuations in Itinerant Electron Magnetism.} (Springer-Verlag, Berlin 
1985). 
%
\bibitem{mp10b} G. Lonzarich and L. Taillefer, J. Phys. C {\bf 18}, 4339 
(1985).  
%
\bibitem{mp11} W. Metzner and D. Vollhardt, Phys. Rev. Lett. {\bf 
62}, 324 
 (1989). 
%
\bibitem{mp12} N. Bl\"umer, J. Wahle, J. Schlipf, K. Held, and D. Vollhardt, 
in preparation. 
%
\bibitem{mp7} J. Hertz, Phys. Rev. B {\bf 14}, 1165 (1976). 
%
\bibitem{mp8} A. J. Millis, Phys.Rev. B {\bf 48}, 7183 (1993). 
%
\bibitem{mp9} T. Vojta, D. Belitz, R. Narayanan, and T. R. Kirkpatrick, 
Europhys.Lett. {\bf 36}, 191 (1996) 
%
\bibitem{mp17} T. Vojta, D. Belitz, R. Narayanan, and T. R. Kirkpatrick, 
Z. Phys. B {\bf 103}, 451 (1997). 
%
%
\bibitem{mp16} C. Pfleiderer, G. J. McMullan, S. R. Julian, and 
G. G. Lonzarich, Phys. Rev. B {\bf 55}, 8330 (1997). 
%
\bibitem{mp19} G. Gumbs and A. Griffin Phys. Rev. B {\bf 13}, 5054 (1976) 
%
\bibitem{mp18} D. Vollhard, N. Bl\"umer, K. Held, M. Kollar, J. 
Schlipf, and M. Ulmke, Z. Phys. B {\bf 103}, 283 (1997). 
%
\bibitem{mp13} D. Schmeltzer, Phys. Rev. B {\bf 43}, 8650 (1991). 
%
\bibitem{mp14} The present discussion is correct if 
the renormalized exchange coupling constant is 
 positive. To this end, it is 
sufficient the microscopic parameters 
 to satisfy the inequality $\frac 
{4t^2}{U} < J$.  
%
\bibitem{mp20} D. P. Arovas, and A. Auerbach, Phys. Rev. B {\bf 38}, 1316 
(1988) 
%
\bibitem{mp21} N.Karchev, Phys. Rev B {\bf 55}, 6372 (1997) 
\end{references}
\end{document}